\documentclass{elsart}

\usepackage{harvard}

\usepackage{graphicx}
\usepackage{amssymb}

\def\url#1{{\ttfamily\def\/{/\discretionary{}{}{}}#1}}

\begin{document}

\begin{frontmatter}
\title{The lives of FR I radio galaxies}

% use the thanksref command within \title, \author or \address for footnotes:
% \title{\thanksref{label1}}
% \thanks[label1]{}
% \author{\thanksref{label2}}
% \thanks[label2]{}
% \address{\thanksref{label3}}
% \thanks[label3]{}
% including your email address
% \address{\thanksref{email}}
% \thanks[email]{E-mail: }

\author[Parma]{P. Parma\thanksref{pp}},
\author[Murgia]{M. Murgia\thanksref{mm}},
\author[de Ruiter]{H.R de Ruiter\thanksref{hr}},
\author[Fanti]{R. Fanti\thanksref{rf}}

\thanks[pp]{E-mail: parma@astbo1.bo.cnr.it}
\thanks[mm]{E-mail: murgia@astbo1.bo.cnr.it}
\thanks[hr]{E-mail: deruiter@kennet.bo.astro.it}
\thanks[rf]{E-mail: rfanti@astbo1.bo.cnr.it}

\address[Parma]{Istituto di Radioastronomia, via Gobetti 101, I-40129 Bologna, Italy}
\address[Murgia]{Istituto di Radioastronomia, via Gobetti 101, I-40129 Bologna,
Italy and Dipartimento di Astronomia, Universita' di Bologna, via Ranzani 1, I-40127, Italy}
\address[de Ruiter]{Istituto di Radioastronomia, via Gobetti 101, I-40129 Bologna, Italy and Osservatorio Astronomico, via Ranzani 1, I-40127, Italy}
\address[Fanti]{Istituto di Radioastronomia, via Gobetti 101, I-40129 Bologna,
Italy and Dipartimento di Fisica, Universita' di Bologna, via Irnerio 46, I-40126, Italy}

\begin{abstract}
After a brief introduction to the morphological properties
of FRI radio sources, we discuss the possibility
 that FRI jets are relativistic at their bases and
decelerate quickly to non-relativistic velocities.

From two-frequency data we determine spectral index distributions and
consequently the ages of FRI sources. We show that in the large
majority of cases synchrotron theory provides unambiguous and plausible
 answers; in a
few objects re-acceleration of electrons may be needed. The derived ages
are of the order $10^7-10^8$ years, $\sim 2-4$ times larger than the ages
inferred from dynamical arguments and a factor $\sim 5-10$ larger
than the ages of FRII sources. The linear sizes of FRI and FRII sources
make it unlikely that many FRII's evolve into FRI's.

A brief discussion is given of the possibility that radio sources
go through different cycles of activity.

\end{abstract}

%\begin{keyword}
% keywords here, in the form keyword \sep keyword
% PACS code here, in the form \PACS code \sep code
%\PACS 
%\end{keyword}
\end{frontmatter}

% main text
\section{FR I radio sources: morphologies and jet velocities}
\label{FR I}

Low and high luminosity radio sources appear to be very different, in fact so much
so, that the Fanaroff and Riley classification not only stresses the
morphological differences but also suggests fundamental physical differences.

\begin{figure}
\begin{center}
\includegraphics*[width=10cm]{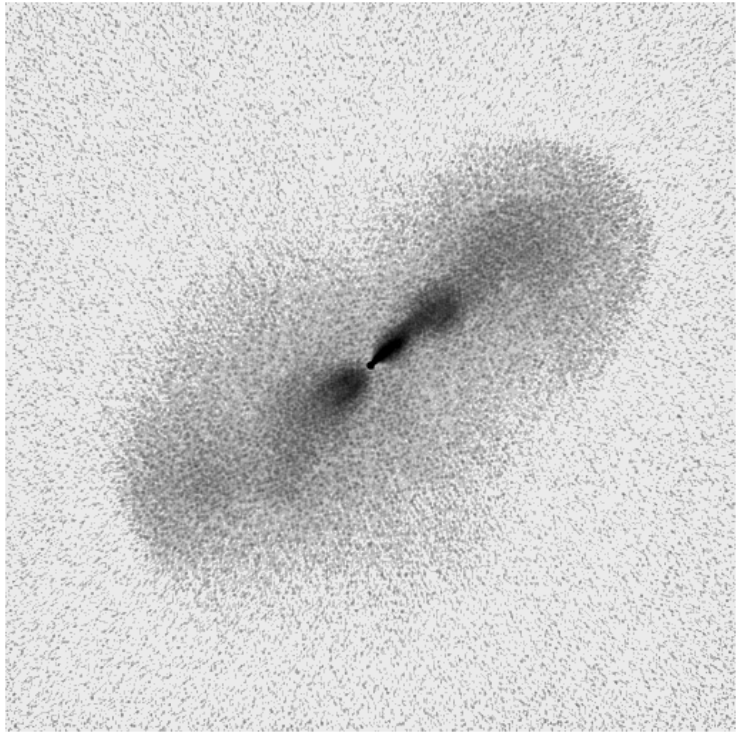}
\end{center}
\caption{B2~0206+35, a typical FRI source.}
\label{0206}
\end{figure}

FRI sources are much more common in the universe than FRIIs and show much
more diversity. While FRIIs are quite homogeneous in morphology the FRI 
class is a kind of  `` mixed bag'' where 
one finds very different sources, which have, apart from the similar radio
power, only one feature in common: hot-spots  
at  the outer end of the lobes are never seen.  3C31 with its twin jet ending in
characteristc plumes has often been  cited as a kind of prototype  FRI
source, but in reality double lobed sources with jets are the rule
 ( Fig. \ref{0206} ):
In the B2 sample of low luminosity radio galaxies 3\% 
are of the 3C 31 type, 50 \% are double lobed with 
(single or double)
jets  and the 
rest is a mixture of tailed sources (NATs and WATs)
and naked jets.

\begin{figure}
\begin{center}
\includegraphics*[width=10cm]{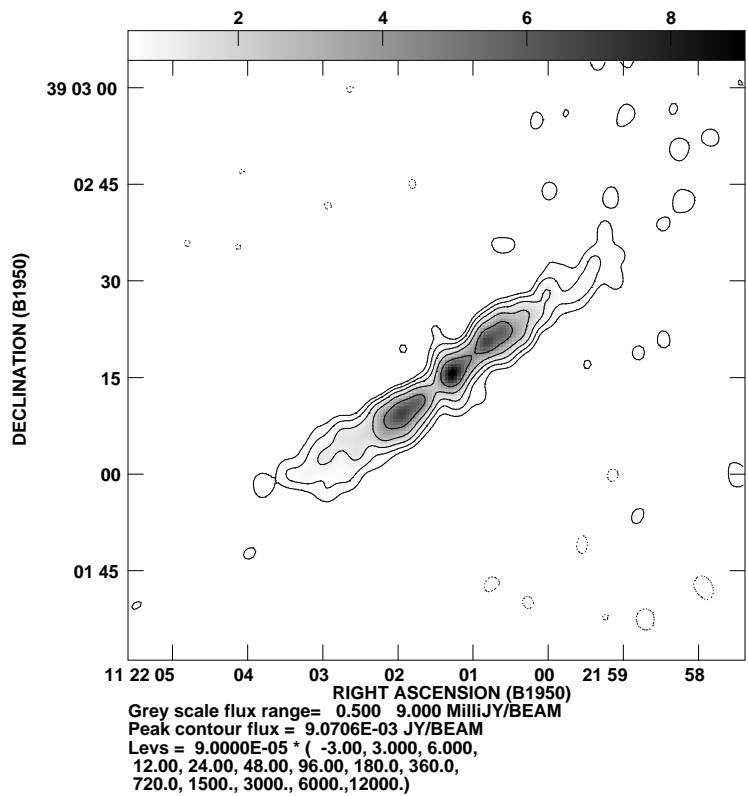}
\end{center}
\caption{B2~1122+39, a ``naked'' jet source.}
\label{1122}
\end{figure}

\begin{figure}
\begin{center}
\includegraphics*[width=10cm]{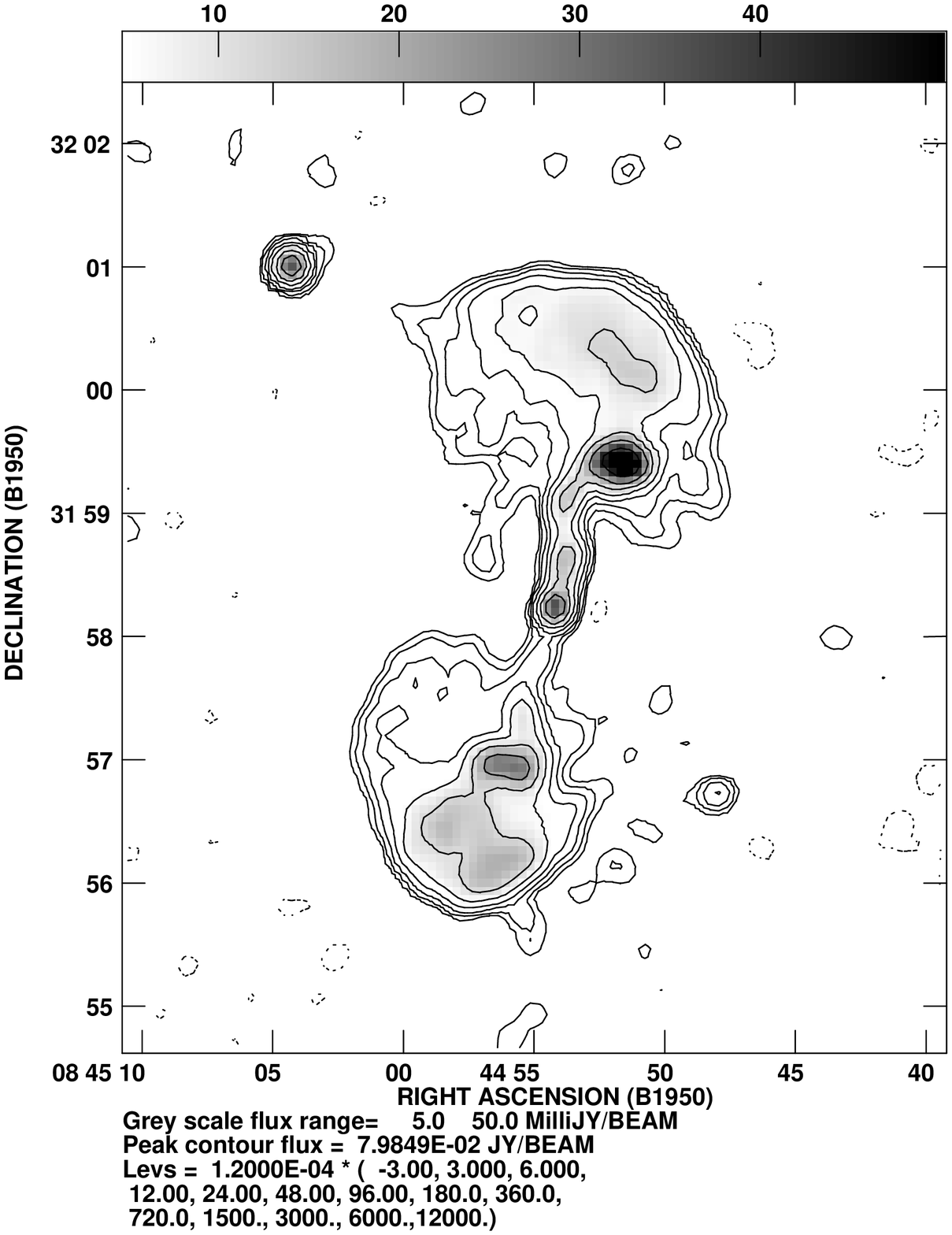}
\end{center}
\caption{B2~0844+31, a FRI - FRII source.}
\label{0844}
\end{figure}

 Naked jets 
(ser e.g. Fig. \ref{1122})
 occur preferable at the low end of the power range 
covered by the B2 sample ($ < 10^{23}$W~Hz$^{-1}$),
 while at the upper end
($ \sim 10^{25}$~WHz$^{-1}$)
we find sources with an overall FR I structure but with one-sided jets and
 hot spots in the middle of the lobes (Fig. \ref{0844}).

A characteristic feature of FR I sources is the presence of prominent twin jets
that are usually very symmetrical at large scales; however the jets tend to be very 
asymmetric close to the nucleus, at the jet base.
This asymmetry is more pronounced in the stronger sources: the
jets tend to become one-sided, more collimated, while brightness knots
appear as we go to higher powers. 

\begin{figure}
% center on page
\begin{center}
% angle=-90 causes the image to be rotated counter-clockwise 90 degrees
\includegraphics*[width=10cm]{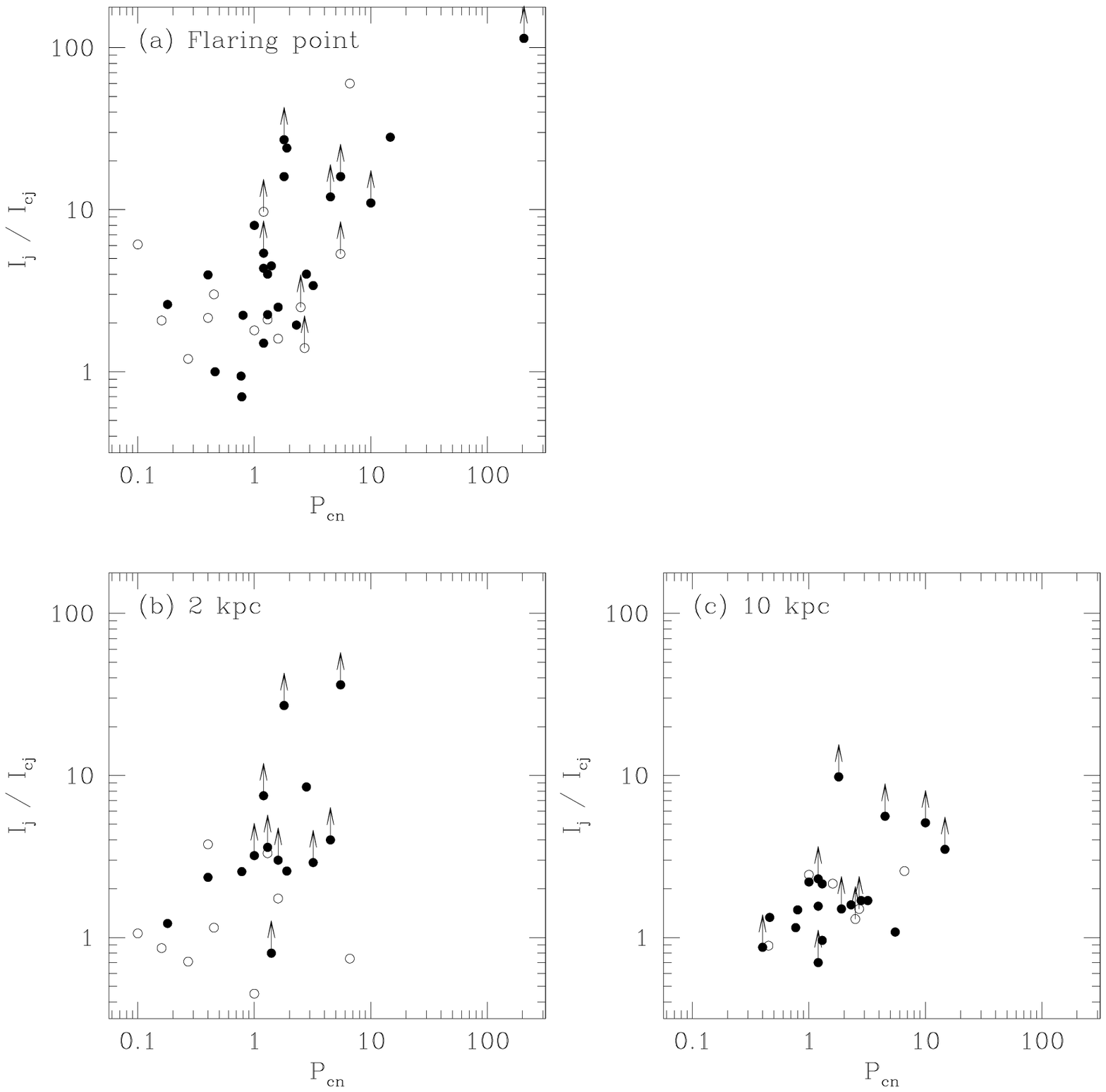}
\end{center}
\caption{Plots of brightness ratio against
normalized core power.  
Open circles: $P_t <$ 10$^{24}$ WHz$^{-1}$; filled 
circles: $P_t \geq$ 10$^{24}$ WHz$^{-1}$. (a) Flaring point, (b) 2~kpc from core,
(c) 10~kpc from core. }
% \caption{This is some random figure of a jet.  From \citeasnoun{Reid89}.}
\label{jet}
\end{figure}

A recent study done by Laing et al. (1999), of the side-to-side
asymmetries on kiloparsec scales 
in the jets of FRI radio galaxies selected from the B2 sample, shows
that differences between the jets at a given distance from
the nucleus can be interpreted as effects of Doppler beaming on intrinsically
symmetrical flows.
 The observed  asymmetries  are consistent with the hypothesis that
the jet velocities close to the core are (mildly) relativistic, and that they
decelerate further out. The length of the deceleration region depends
on total power of the source, and ranges from  $\sim 2$ kpc  for P $ < 10^{24}$~WHz$^{-1}$
 to $\sim$ 10 kpc for the stronger sources.
This point is illustrated in Fig. \ref{jet} which shows the jet to counter-jet brightness ratio as function of 
normalized core power P$_{cn}$ at different locations from the core.
P$_{cn}$ is an independent orientation
indicator which we expect to correlate with the brightness ratio where the
flow is relativistic.  This is indeed the case for the ratio at the flaring
point for both powerful and weak sources (fig. \ref{jet}a). At 2 kpc from the core
the correlation still exists for powerful sources
(fig. \ref{jet} b) but at 10 kpc the correlation has disappeared in both categories of sources (fig. \ref{jet}c).

\begin{figure}
\begin{center}
\includegraphics*[width=10cm]{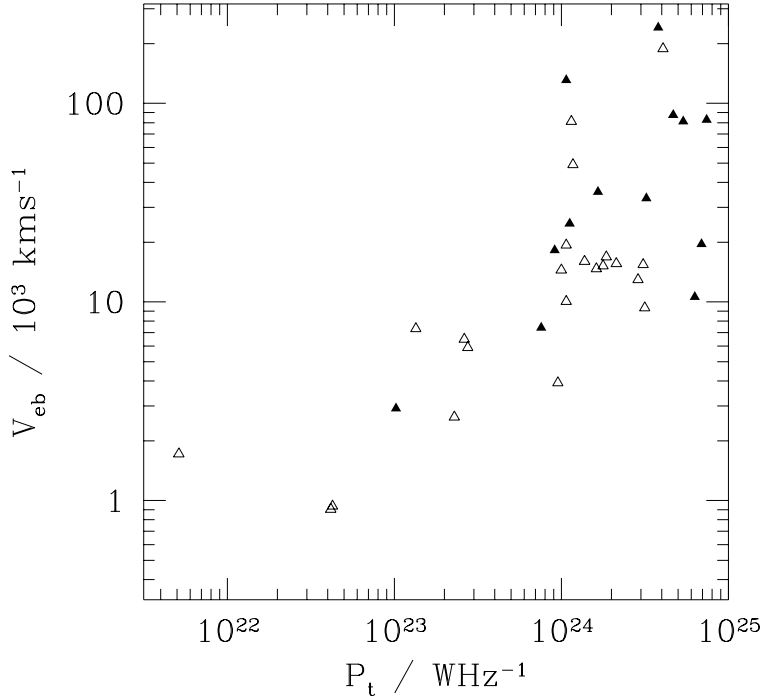}
\end{center}
\caption{A plot of the velocity at 10~kpc
estimated from energy balance against power.  Filled triangles: sources which
have $I_j/I_{cj} >$ 1.5 at 10~kpc from the nucleus (or the end of the jet,
if closer). Open triangles: sources with $I_j/I_{cj} \leq$ 1.5.
 }
\label{vel}
\end{figure}

Also the residual velocity, after the deceleration has taken place, depends on total power.
The jet velocity can be derived by estimating the energy flux through the
jet, $F_E$, from the corresponding lobe luminosity (Bicknell 1986, 1994).
 The resulting velocity
estimates, $v_{eb}$, are plotted against total power in Fig. \ref{vel}.  
The median $v_{eb}$ increases with $P_t$, from $\sim$10$^3$~kms$^{-1}$ ($\beta$
= 0.03) at $P_t$ = 10$^{22}$~WHz$^{-1}$ to $\sim$3 $\times$ 10$^{4}$~kms$^{-1}$
($\beta$ = 0.1) at $P_t$ = 10$^{25}$~WHz$^{-1}$ and the corresponding energy
fluxes are in the range 10$^{33}$ -- 3 $\times$ 10$^{36}$~W.  
The median relation between $v_{eb}$ and $P_t$ is in agreement with
estimates from jet asymmetry in that Doppler beaming is predicted to be
significant at 10~kpc  only for the most powerful sources.

\section{The ages of FR I radio sources}
\label{age}
\subsection{Introduction}

The determination of ages of extragalactic radio sources is one of the key
points for any theoretical model that wants to explain the origin and
evolution of radio galaxies.

One of the methods for determining source ages 
is the study of the radio spectra, for which synchrotron 
theory predicts a frequency break due to the radiative energy losses, which 
drifts in time. According to various source evolution models, relativistic 
electrons in different regions of the source are deposited there at 
different times, so that the break frequency effectively is
a clock that indicates the time elapsed since their production.

For a limited number of objects, detailed studies of the radio spectra across 
the emitting regions  have produced break frequency  and 
source age maps (see e.g. Alexander 1987; Carilli et al. 1991; 
Feretti  et al. 1998).
However, in the majority of these studies the radio spectra across a source 
are based on only one pair of frequencies, so that the break frequency cannot
be seen directly from the data. Nevertheless, with some additional assumptions
which find their justification from the results obtained from the well
studied objects, the break frequency can be estimated with reasonable
 accuracy. 
In fact, the data available in the literature which
concern a large number of powerful 3C sources 
are analyzed
 in this way
(see e.g. Alexander \& Leahy 1987; Leahy et al. 1989; Liu et al. 1992).

Recently
a two frequency spectral study has been done of a 
representative subsample of  B2 radio galaxies (Parma et al., 1999);
in which  the radiative ageing of the relativistic electrons
 caused by synchrotron
and inverse Compton (I.C.) energy losses is discussed.
This study is complementary to 
those dealing with powerful 3CR radio sources: the B2 radio sources are
less powerful by about two orders of magnitude (we use 
$\rm H_0 = 100$~km~s$^{-1}$Mpc$^{-1}$).

\subsection{The data}

The B2 sample of radio sources, which  
contains about a hundred objects, was obtained from identification of B2 radio
sources with bright elliptical galaxies (see Colla et al. 1975; Fanti et al.
1978).  
Because  of the selection criteria, the sample is dominated by
radio galaxies with a power typically  between $10^{23}$ and $10^{25}$ WHz$^{-1}$
at 1.4~GHz. 
Most are FRI sources.

The  sample has been extensively studied with different VLA configurations
at 1.4 GHz (see references in Fanti et al. 1987) and  more recently at 5 GHz
Morganti et al. (1997). 

From the
original B2
  sample only sources with 
 a ratio of overall source size to beam size $\ge $ 10 were selected,
for a total of 32 sources.
For each source the average of the two frequency spectral index 
$\alpha_{5.0}^{1.4}$ was computed
in  slices perpendicular to the source axis
in order to study
 the variation of the spectral index along the source major axis.
 Only data with  a signal to noise ratio  $>$ 5 were used.
The slices are one beam across, so that the data points are practically 
independent. Regions containing radio jets and hot spots were  excluded.
\\
Spectral index errors along the profiles are mostly 
determined by the map noise and
in a minority of cases by dynamic range. They are typically $\sim$ 0.05, but
can be as high as $\geq $ 0.1 in the faintest regions considered.
Unfortunately the cut-off imposed on brightness at 5 GHz introduces a bias
against the inclusion of regions with very steep spectra.\\
From a literature search   additional information 
on the spectral
index distribution
were   recovered
 of 15 
more
objects in the B2 radio galaxy sample.
 These data
have been re-analyzed
 in order to ensure 
as much as possible homogeneity.

\begin{figure}
\begin{center}
\includegraphics*[width=10cm]{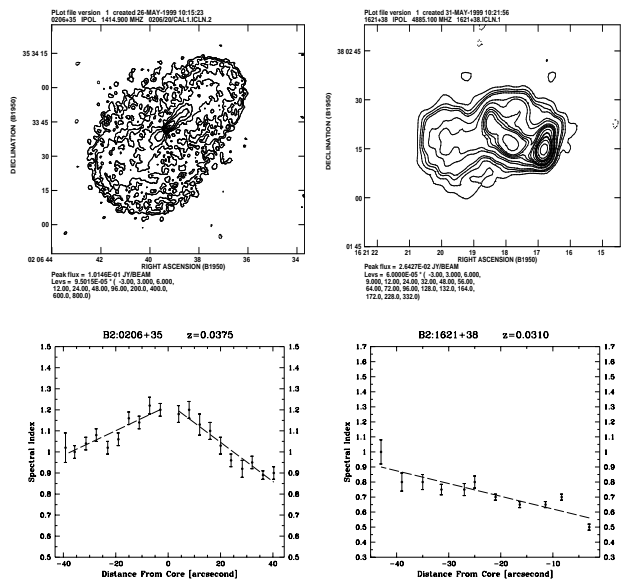}
\end{center}
\caption{Left: spectral type~2; right: spectral type~1}
\label{sptype}
\end{figure}

In most objects the spectral index clearly varies 
with distance from the core. 
The spectrum either steepens from the lobe outer edge inward (hereafter
referred as ``spectral type 2''),
or from the core outward (``spectral type 1''). 
Only a minority of sources do not 
show any significant spectral index trend.
In Fig.~\ref{sptype}  we show the behaviour of the spectral index  for 
type 1 and type 2 sources separately
This double behaviour had already been pointed out some time ago by J{\"a}gers (1981).
In the  ``spectral type 2'' sources the older electrons are found closer to the core. 
This class contains essentially  
sources with double lobed morphology, of both FR~I and FR~II type.
Their spectral behaviour is that 
expected on the basis of standard source models where the radiating particles 
are deposited at different times at the end of an advancing beam and remain
in that position or flow back at some speed toward the core. In this model 
the closer to the core the older they are. 
\\

In the ``spectral type 1'' the spectral index steepens away from the core.
This class is known to contain ``3C 31 like'' objects, WATs, and  NATs 
(see, e.g., J{\"a}gers, 1981).
These are the type of objects for which \cite{ksr97} have found
two distinct components, a flat spectrum jet surrounded by a steeper
spectrum sheath

\subsection{The models and the assumptions}

The spectral trends along the sources are interpreted  in 
terms of radiative ageing of the relativistic electrons by synchrotron and
I.C. processes.

Theoretical synchrotron-loss spectra have 
been computed numerically (Murgia 1996) from the  synchrotron 
formulae (e.g., Pacholczyk 1970). 
It is assumed that the synchrotron and 
I.C. losses dominate and that expansion losses and re-acceleration 
processes can be neglected

Two models are generally considered:  i) the Jaffe-Perola
(JP) model (Jaffe \& Perola 1974), in which the time scale for continuous 
isotropization of the 
electrons is assumed to be much shorter than the radiative time-scale;  
ii) the Kardashev-Pacholczyk (KP) model (Kardashev 1962) in which each 
electron maintains its original pitch angle.
 The JP model has been used since the 
I.C. energy losses due to 
the microwave background radiation are as important
as the synchrotron losses, and in the former random
orientations are expected  between electrons and photons.

\begin{figure}
\begin{center}
\includegraphics*[width=10cm]{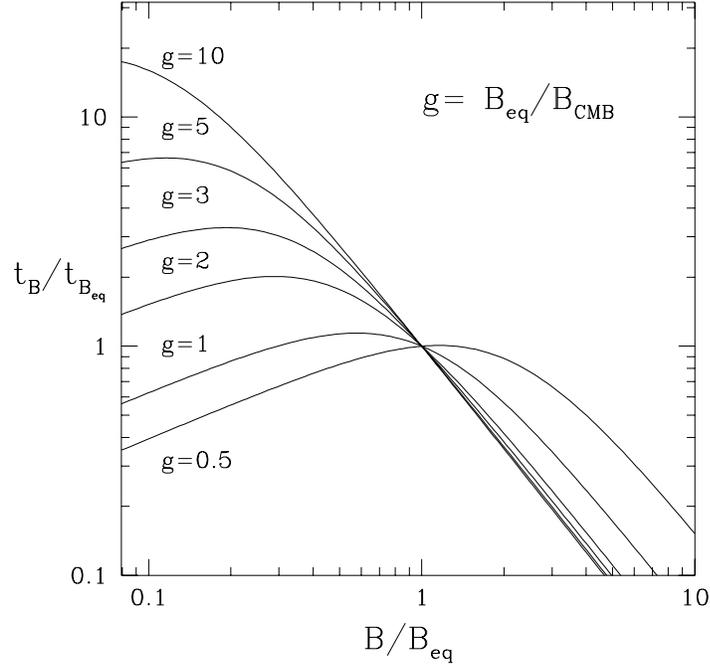}
\end{center}
\caption{The radiative lifetime as a function of the ratio $B/B_{eq}$. }
\label{b}
\end{figure}

The two frequency spectral index $\alpha ^{\nu_1}_{\nu_2}$ allows 
the computation of
 the break frequency, $\nu_{br}$, in various regions of the source, 
using the synchrotron-loss spectrum for the JP  model.
The break frequency is used to determine a spectral age, based on the
formula:
$$t_s = 1.61 \times 10^3 \frac {B^{0.5}}{(B^2+B_{CMB}^2)(\nu_{br} (1+z))^{0.5}}$$
where the synchrotron age $t_s$ is in Myrs, the magnetic field B is in 
$\mu G$,
the break frequency $\nu_{br}$ in GHz and $B_{CMB} = 3.2 \times (1+z)^2$
$\mu G$ is the equivalent magnetic field of the cosmic microwave background
radiation.
It is assumed that the magnetic field strength is uniform across the source 
and has remained constant over the source life.
The magnetic field is computed using the ``minimum energy assumption''.  
 Equality  of protons and electrons energy 
was assumed, a filling factor 
of unity, a radio spectrum ranging from 10 MHz to 100 GHz and an ellipsoidal 
geometry.
/break/hfil
For most of the objects of the sample the computed magnetic field is within a factor 2 of 
$B_{CMB}$. This 
ensures that the synchrotron ages $t_s$ are relatively independent of moderate
deviations from the minimum energy conditions.
\\
Fig. \ref{b} shows how the radiative life time depends on the ratio 
$B/B_{eq}$. For $ 0.5 < B_{eq}/B_{CMB}
< 2$, deviations from equipartition have a small effect on the computed 
lifetime if $B \le 2 \cdot B_{eq}$.

\subsection{ Spectral analysis}

\begin{figure}
\begin{center}
\includegraphics*[width=10cm]{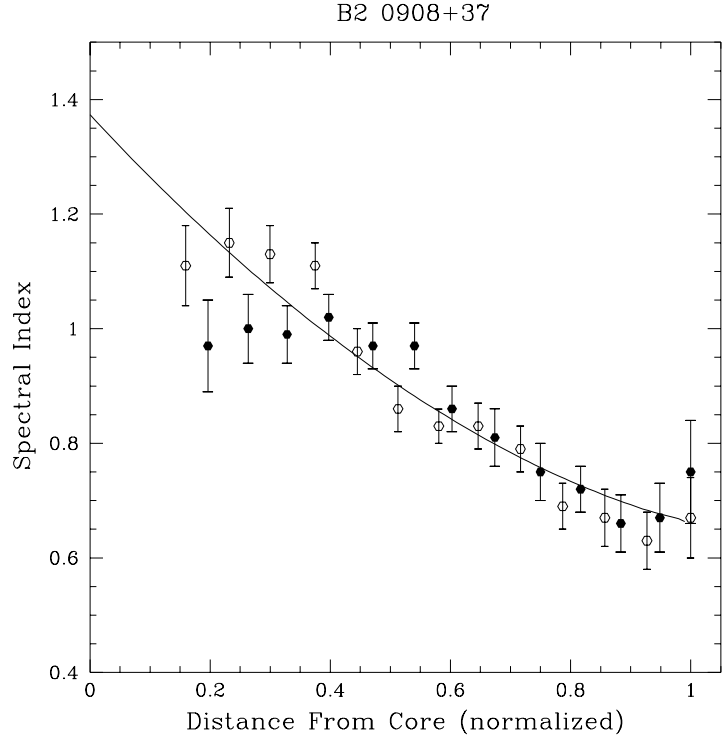}
\end{center}
\caption{  The distances are normalized to the maximum extent of a lobe.
 The full line is 
the best fit of the radiative model. }
\label{fold}
\end{figure}

For the majority of the sources the quality of spectral information is comparable 
in the two lobes, and the spectral trends are rather similar when the
errors are taken into account. Therefore the two lobes
were analyzed
together, instead of considering them separately. 
 This was done by folding the spectral profiles of the two lobes 
onto each other, after having normalized the coordinates along the lobe axis to the 
maximum lobe extent (Fig. \ref{fold}).

 The folded spectral index plots along the source axis 
were fitted with a relationship of the type:
$$\nu_{br} \propto x^{-2}$$
\noindent
where $x$ is the normalized distance from the outer lobe edge or from the 
core according
to the observed steepening trend. 
The above law is expected on the
basis of the relation between radiative age and break frequency, if $x$ is
proportional to time, i.e. assuming a constant expansion speed. From it 
 a best estimate of $\alpha_o$
was obtained, i.e. the spectral  
index close to the outer edge of the lobe or close to the core, according to the
observed spectral profile. This, with a few exceptions, is
assumed to be equal to the injection spectral index, $\alpha_{inj}$.

The mean value of $\alpha_o$ is $ \sim 0.65$,
with a dispersion $\sim 0.1$.
In this way,
 from the folded spectral profiles,
we derived the variation 
of the break frequency as a function of position, averaged 
over the two lobes.
Extrapolating it to the inner or outer part of the source, it is possible to
 obtain the
``minimum break frequency'', $\nu_{br-min}$, namely the break frequency of the
oldest electrons, that we use for determination 
of the source age.
For the large majority of the sources the extrapolation introduces only minor 
additional uncertainties. 
The computed values of $\nu_{br-min}$  range from a few GHz up to several
tens of GHz.

Considering the errors in the spectral index ($\ge 0.05$) and the 
uncertainty in $\alpha_{inj}$ ($\le 0.1$), the uncertainty 
in $\nu_{br-min}$ can be quantified as:
$$\Delta \nu_{br}/\nu_{br} \le 0.08 \cdot (\nu_{br}(GHz))^{0.5}.$$
This implies that only values of $\nu_{br-min}$ $\leq$ 30 GHz 
are significant, which is hardly surprising given that our highest frequency 
is a factor six lower.
The errors on $t_s$ , for a given  magnetic field,
are $\sim$ 15 \%, for break frequencies of a few GHz, increasing up  
to $\geq$ 40 \% for break frequencies $\sim$ 30 GHz.
Had we used the KP model, $t_s$ would be shorter 
by $\sim$ 10 $\%$.

The radiative lifetimes are in the range of $10^7 - 10^8$ years.
 The possibility of 
missing  source areas with steeper spectra
cannot be excluded. Therefore the $t_s$ given are
lower limits on the source ages.

The distributions of radiative ages, $t_s$, for the two spectral classes 
are very similar.

\subsection{Is there particle re-acceleration?}

 The fits 
to the data in general are good, in the sense that they 
describe well the overall trend of $\alpha_{5.0}^{1.4}$ versus $x$. 

There are however a few objects were the break frequency does not always 
seem to increase
 with $x$ as expected from the simple synchrotron/I.C. model.

\begin{figure}
\begin{center}
\includegraphics*[width=10cm]{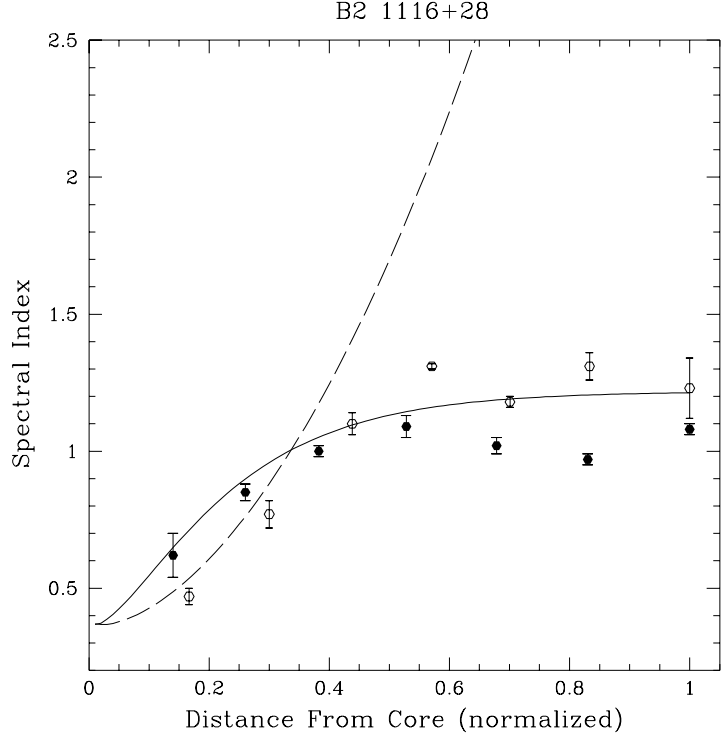}
\end{center}
\caption{Spectral index as a function of distance from the core. The full line
represents the re-acceleration model }
\label{reac}
\end{figure}

The saturation of $\alpha_{5.0}^{1.4}$ to a constant value would 
indicate that the break frequency, after an initial decrease,  
does not decrease anymore. 
A possible explanation of this is that a re-acceleration process is acting,
which compensates in part for the radiative energy losses and causes a freezing 
of the break energy at that value where the radiative and the acceleration 
time scales are equal. Of course, if re-acceleration processes are working, the
ages estimated  would be  underestimated.
Expressing the re-acceleration process as:

\smallskip
$dE/dt = E/ \tau_a,$
\smallskip

where E is the particle energy, and $\tau_a$ is the  
acceleration time scale, the break frequency, as a function of time
(see Kardashev, 1962), is given by:
$$\nu_{br} \propto \frac{B}{(B^2+B_{CMB}^2)^2} \frac{1}{(1-e^{-t/ \tau _a})^2 ~ 
\tau_a^2}
\propto  \frac {\rho^2}{(1 - e^{- \rho ~x})^2},$$
\noindent
where $\rho$ is the ratio of the source age to the re-acceleration 
time scale $\tau _a$ and x is the normalized distance from the outer lobe 
edge or from the core according to the observed steepening trend.
\\
The above law naturally leads to a saturation of $\nu_{br}$ when 
$t \ge \tau _a$.
\\
We have re-fitted all the spectral profiles with the above formula and derived 
a value of $\rho$ for each source and corresponding source age, 
$t_s'$, which is related to the previous one ( $t_s$) by:
$$t_s' = \frac{t_s \rho}{1- e^{- \rho}}$$
For at least 70\%  of the objects the results are not much different from the 
purely radiative model. For those we find  $0 \leq \rho \leq 2$ 
(the median value
is 1.1) and the 
lifetimes are modified by less than a factor 2. Time scales for re-acceleration
must be typically $\geq 4 \cdot 10^7$ years.

However for a few  sources (0844+31, 1116+28, 1521+28, 1528+29)  
the fit improves, while
$\rho$ $>$ 5 suggests  that re-acceleration may be present 
(Fig. \ref{reac}). 
\\
Their ages should then be raised by factors from $\sim$ 6 to $\sim$ 
10 with respect to the ``radiation only'' model.
These  objects are
among those  in which the spectrum steepens away from the core. 
\\     
 There are no significant negative values for $\rho$, which
indicates either that expansion losses are negligible or that they are
well compensated by re-acceleration processes.

\subsection{Discussion}

\subsubsection{Spectral ages and source sizes}

\begin{figure}
\begin{center}
\includegraphics*[width=10cm]{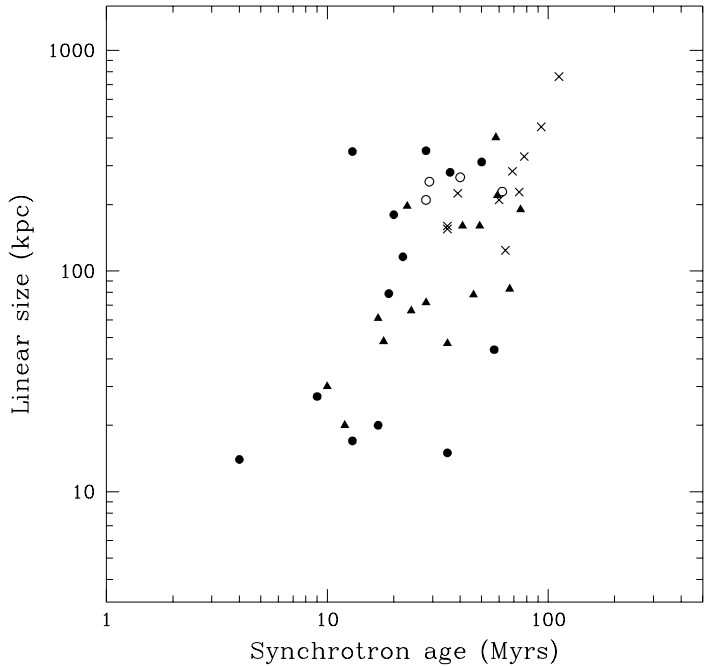}
\end{center}
\caption{Linear size as a function of spectral age of the source.
The crosses represent the sources taken from the literature, the dots type 1
spectra, 
 the open circles referring to four sources  whose radiative lifetimes could be definitely longer.
The triangles represent type 2 spectra. }
\label{age-size}
\end{figure}

It  has been investigated whether there is a relationship between  radiative age and
source size. 
Indeed a significant correlation is present
as shown in Fig. \ref{age-size}.
A linear fit between the logarithms of age and linear size gives:
$$LS \propto {t_{s}}^{0.97},$$
\noindent
where the uncertainty in the exponent is 0.17. 

One might think that the correlation between linear
sizes  and the radiative ages shown  could be a partial consequence
of the equipartition assumption. In fact, $B_{\rm eq} \propto {\rm LS}^{-6/7}
$ implies that ${\rm LS} \propto \tau_{\rm syn}^{7/9}$.
In our case the computed magnetic field is within a factor 2 of $B_{CMB}$,
therefore the synchrotron ages are relatively independent from the minimum
energy conditions.

The advance speed of the lobes, deduced from the synchrotron ages, are in the 
range of $0.5 - 5 \cdot 10^3$~km/sec. There is no difference between the 
two spectral classes.
                            
\subsubsection{A comparison between spectral and dynamical ages}

 The synchrotron ages, $t_s$,
have been compared
 with the dynamical ages, $t_d$,
evaluated from simple ram-pressure arguments.
The expansion velocity of the lobes is given by the relation:
$$v_{exp} \sim \left[\frac{\Pi}{A}~ \frac{1}{m_p  n_e}\right]^{0.5}$$
where $\Pi$ is the jet thrust and $A$ is the size of the area over which 
it is discharged.

The most simplistic assumption is to identify the $\Pi/A$ ratio 
with the hot spot pressure.
   The front surface minimum pressures of the lobes, 
$p_{eq,f}$, as been used as the appropriate quantity 
for $\Pi/A$ (Williams, quoted by Carilli et al., 1991).

The values 
found are typically $\> 4$  times the average minimum lobe pressure.

It is  assumed that the jet thrust is constant over the source 
life time and that the source grows in a self-similar way, such that  
$p_{eq,f}  \propto R_{kpc}^{-2}$. 
Finally  a run of the external density $n_e = n_o \times 
R_{kpc}^{-\beta}$
is assumed, with $n_o \sim 0.5~$ cm$^{-3}$
and $1.5 \le \beta \le 2$,  according to Canizares et al. (1987).
Under these assumptions there would be only a  slight dependence, if any, of 
$v_{exp}$ on source size, depending on the value of $\beta$.

\begin{figure}
\begin{center}
\includegraphics*[width=10cm]{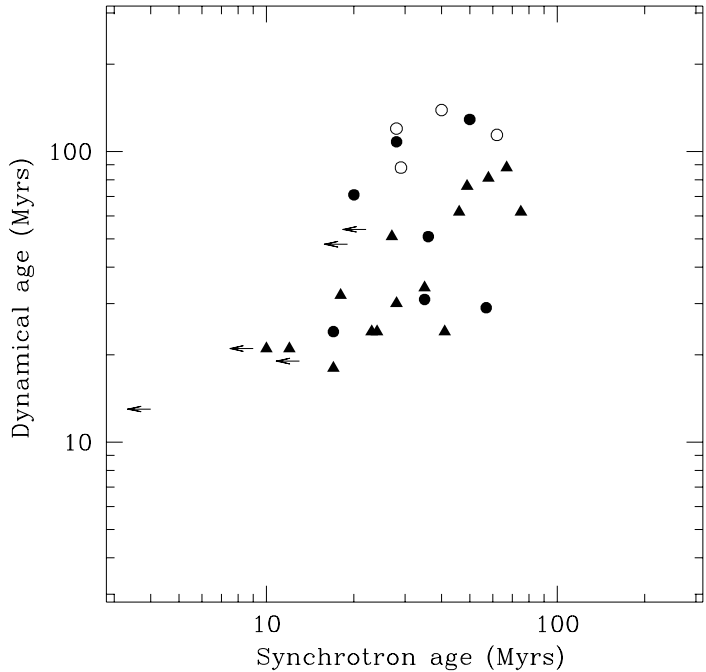}
\end{center}
\caption{Dynamical age vs  spectral age of the source for $\beta = 2.0$.
The dots represent sources with spectral type 1, 
the open circles referring to four sources  whose radiative lifetimes could be definitely longer. Sources with spectra of type 2 are denoted by triangles.
The arrows refer to sources for which the $t_s$ are upper 
limits }
\label{tdyn}
\end{figure}

The dynamical and radiative ages are correlated, but the dynamical ages
in general are larger by a factor $\sim 4$ for $\beta = ~1.5$ and 
$\sim ~2$ for $\beta = 2$ (Fig. \ref{tdyn}). 

These discrepancies are believed to be not very serious for at least three
reasons. 

First of all, the  one-dimensional ram-pressure balance may not be 
realistic and discrepancies of a factor 2 or so are well possible, as 
mentioned above.

Second, the assumed central density  may be a bit too high.
A value around 0.1 cm $^{-3}$, which is not excluded by the X-ray data, 
would bring, for $\beta~ = ~2$, the dynamical ages into closer agreement with
the spectral ones. 

Third, the ram-pressure dynamical 
ages depend on the minimum energy assumption. It has been shown
that radio galaxies with luminosities like in the present sample usually
have internal pressures larger than the minimum ones by a factor $\geq 5$
(Morganti et al. 1988; Feretti et al. 1992).
If  the magnetic field is weaker than the one corresponding
to minimum energy conditions by a factor $\geq$ 4, the internal 
pressure increases 
by a factor $\geq$ 4 and the dynamical lifetime decreases by a factor 
$\sim$ 2.
Anyway, the radiative lifetimes would change little.
Likely  both explanations may play a role in bringing $t_s$ and $t_d$ to
a closer agreement.
\\
Finally,  the possibility of re-acceleration processes is 
another factor which goes in the direction of bringing the radiative 
and the dynamical timescales closer.

\subsubsection{Spectral ages and radio power: a comparison with the 3CR 
radio galaxies}

 The synchrotron ages derived for the B2 sample have been 
compared with the 
corresponding ones of 3CR sources as found by Alexander \& Leahy (1987),
Leahy et al. (1989) and Liu et al.(1992). When necessary the data were
corrected, to take into account different assumed values for
the Hubble constant.

\begin{figure}
\begin{center}
\includegraphics*[width=10cm]{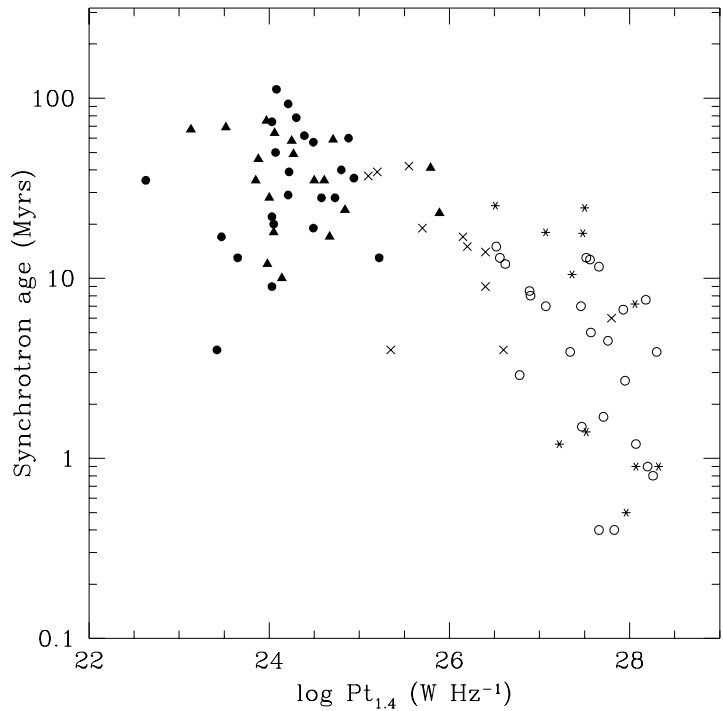}
\end{center}
\caption{Source age versus radio power at 1.4 GHz. The meaning of the symbols is
as follows. Filled circles and triangles: B2 sources with type 1 and type 2 spectra, respectively; 
crosses: 3C galaxies with $z<0.2$; open circles:
3C galaxies with $z>0.2$; asterisks: 3C quasars. }
\label{age-power}
\end{figure}

It appears that the $t_s$ of B2 sources are systematically larger than those
of the 3CR sources by factors 5 - 10 (Fig. \ref{age-power})

There are several possible reasons for these differences:

a)  the radiative ages derived for the 
powerful sources (3CR) are  heavily
dependent on the equipartition magnetic field, $B_{eq}$, which is 
generally significantly larger than $B_{CMB}$. 
For B $\sim  1/4 \cdot B_{eq}$ the differences between the two sample 
would be greatly reduced.

b) Another possibility is that, if there is significant backflow in the 
powerful sources, there is 
mixing of the old and young
 electrons 
in the inner regions, which leads to a
 smoothing of the spectral 
break and an under-estimation of the maximum radiative time scale.

Finally, the effect may be real,
that is
in  low power sources like those of 
B2  sample, the nuclear activity lasts for a longer time. 
One could think of an evolutionary effect, in the sense that low power
sources have evolved from high power sources and are therefore older.
 However,  sources of B2 sample are on average smaller
than powerful 3CR sources (de Ruiter et al., 1990) so that this
explanation is
unlikely in general, even if it cannot be excluded as an explanation in
some individual cases.  

Another possibility is a cosmological effect, since the 
3CR sources are at much larger red-shifts than B2 radio galaxies.

\section{Cycles of nuclear activity}
\label{cycles}

The radiative ages thus computed are much smaller than the ages obtained
from the radio luminosity function.
A long time ago 
Schmidt (1966)
found for powerful radio galaxies
 a harmonic-mean life time of about $10^{9}$ years.
 This life time  is the total time of the nuclear activity,  not to be 
confused with the radiative loss age determined from spectral ageing
arguments.
The length of the active phase is related to the possibile existence of duty 
cycles of nuclear activity. If the nuclear activity is not continuous,
how often the interruptions occur and how long do they last?
A duty cycle can be recognized as such if there is some mechanism that
 preserves the information of past nuclear activity for a long enough time to 
be recognized when a new cycle starts up. In the extended radio sources, such
 a 
mechanism is provided for by the radio lobes. If a new phase of activity
should start before these ``old'' radio lobes have faded, and if this activity
manifests itself by the production of jets, we can in principle recognize 
this as a new,young radio source embedded in a old relic structure.
Of course this requires  that the off time of the central 
activity is shorter than the fade-away time of the old radio lobes.           

\begin{figure}
% center on page
\begin{center}
% angle=-90 causes the image to be rotated counter-clockwise 90 degrees
\includegraphics*[width=8cm, angle=-90]{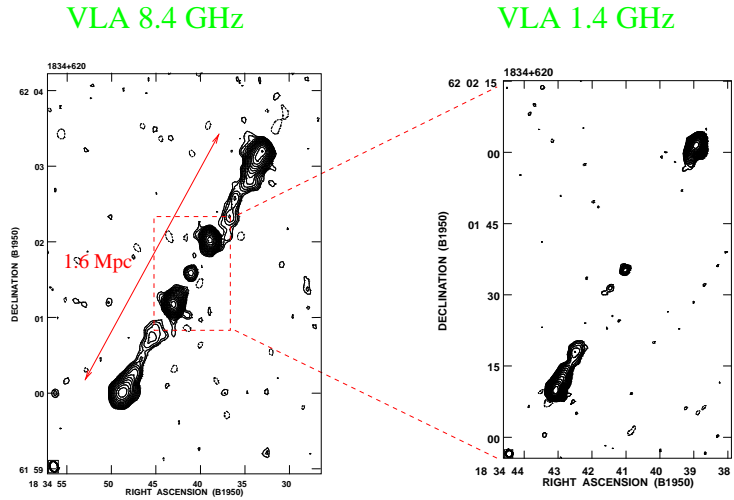}
\end{center}
\caption{The double-double radio galaxy WNB~1834+620. From 
Schoenmakers et al. 1999.}
%\caption{This is some random figure of a jet. From \citeasnoun{schoen99}.}
\label{DDRG}
\end{figure}

It the off time is long enough we have the double - double radio galaxies
(DDRG) (Schoenmakers et al., 1999),
radio sources with four lobes
(Fig. \ref{DDRG}). These radio sources have been discovered
by Arno Schoenmakers in the WENSS survey and they consist of an inner 
double lobed
radio structure as well as a larger outer double lobed structure. The inner
and outer radio sources are well aligned, within 10 degrees. 
Following the definition of Schoenmakers, a DDRG consists of a pair of double
radio sources with a common centre. The two lobes of the inner radio sources 
must  have an extended, edge-brightened radio morphology.
Up to now four DDRG have been found in WENSS survey,
while
  three
more can be founf in the
 literature (3C445, 4C26.35 and 3C219)
The DDRG have Mpc size (the outer components) and the power of the outer lobes
is higher than the power of the inner lobes
 ($ P_{1.4} \sim 10^{25} W Hz^{-1}$).
 
The DDRG provide good evidence for recurrent radio activity in AGN: a large
source is fading away while at the same time a smaller radio source
emerges.

%\clearpage
% The phrase \cite{Bai92} produces (Bailyn 1992).
% In the phrase \citeasnoun{Bai95} Bailyn et al. (1995) appear as a noun.
% Affixes (e.g. Barnes et al. 1976) are produced by the phrase
% \citeaffixed{Barnes et al. 1976}{e.g.}.
% Other options of the harvard package, e.g. \citeyear, are not
% reproduced in New Astronomy.

\end{document}